\def\beq#1{\begin{equation}\label{#1}}
\def\eeq{\end{equation}}
\def\beqa#1{\begin{eqnarray}\label{#1}}
\def\eeqa{\end{eqnarray}}

\def\fun#1#2{\lower3.6pt\vbox{\baselineskip0pt\lineskip.9pt
        \ialign{$\mathsurround=0pt#1\hfill##\hfil$\crcr#2\crcr\sim\crcr}}}
        
\def\xi{{{\bf x}^b}}

\documentclass[twocolumn,aps,showpacs,showkeys,nofootinbib]{revtex4}
\usepackage{epsfig}

\newcommand{\be}{\begin{equation}}
\newcommand{\ee}{\end{equation}}
\newcommand{\ba}{\begin{eqnarray}}
\newcommand{\ea}{\end{eqnarray}}

\begin{document}
\input{epsf.sty}

\title{Model-Independent Distance Measurements from Gamma-Ray Bursts\\
and Constraints on Dark Energy}

\author{Yun~Wang}
\address{Homer L. Dodge Department of Physics \& Astronomy, Univ. of Oklahoma,
                 440 W Brooks St., Norman, OK 73019;
                 email: wang@nhn.ou.edu}

                 \today

\begin{abstract}

Gamma-Ray Bursts (GRB) are the most energetic events in the Universe,
and provide a complementary probe of dark energy by allowing the measurement 
of cosmic expansion history that extends to redshifts greater than 6.
Unlike Type Ia supernovae (SNe Ia), GRBs must be calibrated for each 
cosmological model considered, because of the lack of a nearby sample of 
GRBs for model-independent calibration. For a flat Universe with a cosmological 
constant, we find $\Omega_m=0.25^{+0.12}_{-0.11}$ from 69 GRBs alone.
We show that the current GRB data can be summarized by a set of 
model-independent distance measurements, with negligible loss of information.
We constrain a dark energy equation of state linear in the cosmic scale factor 
using these distance measurements from GRBs, together with the ``Union'' 
compilation of SNe Ia, WMAP five year observations, and the SDSS baryon acoustic 
oscillation scale measurement. We find that a cosmological constant 
is consistent with current data at 68\% confidence level for a flat Universe.
Our results provide a simple and robust method to incorporate GRB data in a 
joint analysis of cosmological data to constrain dark energy.

\end{abstract}



\maketitle


\section{Introduction}

Gamma-ray bursts (GRBs) are the most luminous astrophysical events 
observable today, because they are at cosmological distances \cite{Paczynski95}. 
The duration of a gamma-ray burst is 
typically a few seconds, but can range from a few milliseconds to 
several minutes. The initial burst at gammay-ray wavelengths
is usually followed by a longer-lived afterglow at longer wavelengths 
(X-ray, ultraviolet, optical, infrared, and radio). 
Gamma-ray bursts have been detected by orbiting satellites about two to 
three times per week.
Most observed GRBs appear to be collimated emissions caused by 
the collapse of the core of a rapidly rotating, high-mass star 
into a black hole. A subclass of GRBs (the ``short" bursts) appear 
to originate from a different process, the leading candidate being 
the collision of neutron stars orbiting in a binary system. 
See Ref.\cite{Meszaros06} for a recent review on GRBs.

GRBs can be used as distance indicators \cite{GRBde}, thus
provide a complementary probe of dark 
energy\footnote{Ref.\cite{DEreviews} contains reviews with extensive 
lists of references on dark energy research.}.
The main advantage of GRBs over Type Ia Supernovae (SNe Ia) is that 
they span a much greater redshift range (from low $z$ to $z> 6$). 
The main disadvantage is that GRBs have to be calibrated for 
each cosmological model tested (see for example,
Ref.\cite{Schaefer04}). This is in contrast to SNe Ia, 
where the calibration relations are established using nearby SNe Ia, 
and applied to high $z$ SNe Ia to extract cosmological constraints.
There are no nearby GRBs that can be used for calibration. 
Thus, the GRB data must be fitted simultaneously for calibration
and cosmological parameters. This makes the use of GRBs to
probe cosmology somewhat cumbersome.

In this paper we show that the current GRB data can be summarized
by a set of model-independent distance measurements.
Our results provide an easy, robust, and transparent way to incorporate
GRB data in an analysis of combined cosmological data to
constrain dark energy.

\section{Method}
\subsection{Calibration of GRBs}
\label{sec:cali}

Following \cite{Schaefer07}, we consider five calibration
relations for GRBs.
These relate GRB luminosity, $L$, or the total burst energy in
the gamma rays, $E_\gamma$, to observables of the light curves and/or
spectra: $\tau_{\mathrm{lag}}$ (time lag), $V$ (variability),
$E_{\mathrm{peak}}$ (peak of the $\nu F_\nu$ spectrum), 
and $\tau_{\mathrm{RT}}$ (minimum rise time):
\begin{eqnarray}
  \label{eq:C1}
  \log \left(\frac{L}{1 \; \mathrm{erg} \; \mathrm{s}^{-1}}\right)
  &=& a_1+b_1 \log
  \left[
    \frac{\tau_{\mathrm{lag}}(1+z)^{-1}}{0.1\;\mathrm{s}}
  \right]
  ,
  \\
  \label{eq:C2}
  \log \left(\frac{L}{1 \; \mathrm{erg} \; \mathrm{s}^{-1}}\right)
  &=& a_2+b_2 \log
  \left[
    \frac{V(1+z)}{0.02}
  \right]
  ,
  \\
  \label{eq:C3}
  \log \left(\frac{L}{1 \; \mathrm{erg} \; \mathrm{s}^{-1}}\right)
  &=& a_3+b_3 \log
  \left[
    \frac{E_{\mathrm{peak}} (1+z)}{300\;\mathrm{keV}}
  \right]
  ,
  \\
  \label{eq:C4}
  \log \left(\frac{E_{\gamma}}{1\;\mathrm{erg}}\right)
  &=& a_4+b_4 \log
  \left[
    \frac{E_{\mathrm{peak}} (1+z)}{300\;\mathrm{keV}}
  \right]
  ,
  \\
  \label{eq:C5}
  \log \left(\frac{L}{1 \; \mathrm{erg} \; \mathrm{s}^{-1}}\right)
  &=& a_5+b_5 \log
  \left[
    \frac{\tau_{\mathrm{RT}}(1+z)^{-1}}{0.1\;\mathrm{s}}
  \right]
  .
\end{eqnarray}

Not surprisingly, $E_{\mathrm{peak}}$ carries the most distance information.
The $E_{\mathrm{peak}}$ -- $E_{\gamma}$ relation is the tightest of the 
GRB calibration relations. To be included in this relation, 
the GRB afterglow must have an observed jet break 
in its light curve, and this means that only a fraction of GRBs 
with redshifts can contribute to establishing this relation.  
The variability-luminosity relation has the largest scatter.
The variability is a measure of the sharpness of the pulse structure, 
which is determined by the size of the visible region in the jet.

In order to calibrate GRBs, $L$ and $E_\gamma$ must be related 
to the observed bolometric peak flux, $P_{\mathrm{bolo}}$, and the
bolometric fluence, $S_{\mathrm{bolo}}$:
\ba
L&=&4\pi d_L^2 P_{\mathrm{bolo}}\nonumber\\
E_\gamma&=&E_{\gamma,\mathrm{iso}} F_{\mathrm{beam}}=
4\pi d_L^2 S_{\mathrm{bolo}}(1+z)^{-1} F_{\mathrm{beam}},
\ea
where $E_{\gamma,\mathrm{iso}}$ is the isotropic energy.
Clearly, the calibration of GRBs depend on the cosmological model
through the luminosity distance $d_L(z)$.

The cosmological constraints from GRBs are sensitive to how the
GRBs are calibrated. Calibrating GRBs using Type Ia supernovae (SNe Ia)
gives tighter constraints than calibrating GRBs internally \cite{Liang08}.
In this paper, we choose to calibrate GRBs internally, without
using any external data sets, so that our results can be used to
combine with any other cosmological data sets.

In fitting the five calibration relations, we need to fit a data array
$\{x_i, y_i\}$ with uncertainties $\{\sigma_{x,i}, \sigma_{y,i}\}$, to a 
straight line
\be
y =a + b x
\ee
through the minimization of $\chi^2$ given by \cite{Press07}
\be
\label{eq:chi2fit}
\chi^2= \sum_{i=1}^{N} \frac{ \left(y_i - a - b x_i\right)^2}{
\sigma_{y,i}^2+ b^2 \sigma_{x,i}^2 }.
\ee
It is convenient to define 
\be
x_i^{(\alpha)} \equiv \log\left(x_{0,i}^{(\alpha)}\right),
\ee
thus
\ba
x_{0,i}^{(1)} & =  & \frac{\tau_{\mathrm{lag},i}(1+z)^{-1}}{0.1\;\mathrm{s}} \\
x_{0,i}^{(2)} & =  & \frac{V(1+z)}{0.02}\\
x_{0,i}^{(3)} & = & x_{0,i}^{(4)}= \frac{E_{\mathrm{peak},i} (1+z)}{300\;\mathrm{keV}}\\
x_{0,i}^{(5)} & =  &\frac{\tau_{\mathrm{RT},i}(1+z)^{-1}}{0.1\;\mathrm{s}}
\ea
and
\ba
y_i^{(1)}&=&y_i^{(2)}=y_i^{(3)}=y_i^{(5)}=
\log \left(\frac{L}{1 \; \mathrm{erg} \; \mathrm{s}^{-1}}\right) \nonumber\\
&=&\log(4\pi P_{\mathrm{bolo},i})+ 2\log \overline{d_L},\nonumber\\
y_i^{(4)}&=&\log \left(\frac{E_{\gamma}}{1\;\mathrm{erg}}\right)\nonumber\\
&=& \log\left[\frac{4\pi S_{\mathrm{bolo},i}F_{\mathrm{beam},i}}
{1+z}\right] 
+ 2\log \overline{d_L},
\ea
where we have defined 
\be
\overline{d_L} \equiv (1+z) H_0 r(z)/ c.
\ee
Since the absolute calibration of the GRBs is unknown, the Hubble constant
cannot be derived from GRB data.
Thus we have defined the data arrays $\{y_i\}$ such that 
\be
c/H_0 =9.2503\times 10^{27} h^{-1}\, \mathrm{cm}
\ee
is absorbed into the overall calibration. 

Furthermore, for the $L$ -- $E_{\mathrm{peak}}$ and 
$E_{\mathrm{peak}}$ -- $E_{\gamma}$
relations, the measurement error of $E_{\mathrm{peak}}$ is asymmetric, thus
we need to modify the $\chi^2$ such that
\ba
\sigma_{x,i} &=&\sigma_{x,i}^+, \hskip 1cm \mathrm{if}\,\, (y_i-a)/b \ge x_i; \nonumber\\
\sigma_{x,i} &=&\sigma_{x,i}^-, \hskip 1cm \mathrm{if}\,\, (y_i-a)/b < x_i,
\label{eq:dx asy}
\ea
where $\sigma_{x,i}^+$ and $\sigma_{x,i}^-$ are the $\pm$ measurement
errors.

As noted by Ref.\cite{Schaefer07}, the statistical errors on $\{a_i,b_i\}$
are quite small, but the $\chi^2$'s are very large due to the domination
of systematic errors. Following Ref.\cite{Schaefer07}, we derive the
systematic errors by requiring that $\chi^2 = \nu$ (the degrees of freedom),
and that $\sigma_{tot}^2=\sigma^2_{stat}+\sigma^2_{sys}$.

For illustration on the cosmological parameter dependence of the calibration
of GRBs, Table 1 shows the systematic errors, as well as the constants
$\{a_i,b_i\}$ for the five calibration relations
for $\Omega_m=$0.2, 0.27, 0.4 for a flat Universe with a cosmological constant. 
Note that the $a_i$ in this table 
is smaller than the definition of Ref.\cite{Schaefer07} 
by 2$\log(9.2503\times 10^{27} h^{-1})$.
Note also that the derived systematic errors change by less than 3\% for the
different models. Since systematic errors should be independent 
of the cosmological model, we take the systematic errors for
the $\Omega_m=0.27$ flat $\Lambda$CDM model to be the standard
values in the rest of this paper.

For most of the calibration relations, fitting straight lines using
errors in both coordinates does not give significantly different
results from fitting straight lines using errors in the $y$ coordinates
only. For the $L$ -- $V$ relation, assuming the $\Omega_m=0.27$ flat $\Lambda$CDM model, 
fitting straight lines using errors in the $y$ coordinates only,  we find
$a_2= -3.540\pm 0.003$, $ b_2= 1.649\pm 0.012$, and $\sigma_{sys,2}=0.518$.
The slope and the systematic error are both significantly smaller than
the results shown in Table 1.
Since the $x$ coordinates have significant measurement errors in
all five calibration relations, the latter should be fitted
to straight lines using errors in both $x$ and $y$ coordinates.

\begin{table*}[htb]
\caption{Systematic errors for the five GRB calibration relations}
\begin{center}
\begin{tabular}{|l|l|l|l|}
\hline
 & $\Omega_m=0.27$ & $\Omega_m=0.2$ & $\Omega_m=0.4$ \\
 \hline
$a_1$ & $-3.901\pm$ 0.027 &$-3.848 \pm$ 0.027 &$-3.979 \pm$ 0.026 \\
$b_1$ & $-1.154\pm$ 0.033 &$ -1.167 \pm$ 0.033 &$-1.138 \pm$ 0.032\\
$\sigma_{sys,1}$ & 0.417  & 0.427 & 0.405 \\
 \hline
$a_2$ &$-3.822 \pm$ 0.011 &$-3.781 \pm$ 0.011 & $-3.883 \pm$ 0.011 \\
$b_2$ &3.983 $\pm$ 0.050 & 4.017 $\pm$ 0.05 &3.934 $\pm$ 0.049 \\
$\sigma_{sys,2}$ & 0.927 & 0.931 & 0.922 \\
 \hline
$a_3$ &$-3.821 \pm $ 0.010 &$-3.782 \pm$ 0.010 & $-3.881\pm$0.010\\
$b_3$ &1.830 $\pm$ 0.027 & 1.863 $\pm$ 0.027 &1.787 $\pm $ 0.026 \\
$\sigma_{sys,3}$ & 0.466 & 0.466 & 0.467 \\
 \hline
$a_4$ &$-5.612 \pm$ 0.024 &$-5.574 \pm$ 0.024 & $-5.672 \pm$ 0.024\\
$b_4$ &1.452 $\pm$ 0.059 &1.468 $\pm$ 0.060 & 1.430$\pm$ 0.058\\
$\sigma_{sys,4}$ &0.204 &0.200 & 0.211 \\
 \hline
$a_5$ &$-3.486 \pm$ 0.023 & $-3.431 \pm$ 0.024 & $-3.567 \pm$0.023\\
$b_5$ &$-1.590 \pm $ 0.044 & $-1.617 \pm$ 0.044 & $-1.557\pm $0.042\\
$\sigma_{sys,5}$ & 0.591 & 0.598 & 0.582\\
\hline
\end{tabular}
\end{center}
\end{table*}

Note that we do {\it not} use the calibration relations from Table 
1 (the parameters $a_i$ and $b_i$) when we derive model-independent 
distances from GRBs (see Sec.IIB).
Table 1 is only used to show that the calibration of GRBs is sensitive 
to the assumptions about cosmological parameters, but the systematic
uncertainties of the calibration parameters $a_i$ and $b_i$ are
{\it not} sensitive to the assumptions about cosmological parameters.
Thus, we will derive calibration relations of GRBs for each
set of assumed distances, but we will assume that the systematic
errors on $a_i$ and $b_i$ are given by the $\Omega_m=0.27$ flat 
$\Lambda$CDM model.

\subsection{Model-independent distance Measurements from GRBs}

Following Ref.\cite{Schaefer07}, we weight the five estimators
of distance of GRBs (from the five calibration relations) as follows
\ba
&&\left(\log \overline{d_L}^2\right)_i^{data} = \frac{\sum_{\alpha=1}^5 
\left(\log \overline{d_L}^{2}\right)_i^{(\alpha)} /\sigma_{i,\alpha}^2}
{\sum_{\alpha=1}^5 1/\sigma_{i,\alpha}^2} \\
&&\sigma\left(\log \overline{d_L}^2\right)_i^{data}= 
\left(\sum_{\alpha=1}^5 1/\sigma_{i,\alpha}^2\right)^{-1/2}
\ea
where
\ba
\left(\log \overline{d_L}^2\right)^{(\alpha)}_i&=& a_\alpha+b_\alpha x_i^{(\alpha)}
-\log(4\pi P_{\mathrm{bolo},i}), \nonumber\\
& & \mathrm{for}\,\, \alpha=1,2,3,5 \\
\left(\log \overline{d_L}^2\right)_i^{(4)}&=& a_4+b_4 x_i^{(4)} -
\log\left[\frac{4\pi S_{\mathrm{bolo,}i} F_{\mathrm{beam,}i}}{1+z}\right],\nonumber\\
\ea
and
\ba
\sigma_{i,\alpha}^2 &=& \sigma_{a_\alpha}^2+ \left( \sigma_{b_\alpha} 
x_i^{(\alpha)}\right)^2 + \left( \frac{b_\alpha \sigma(x_{0,i}^{(\alpha)})}
{x_{0,i}^{(\alpha)}\ln 10} \right)^2  \nonumber\\
& & + \left( \frac{\sigma(P_{\mathrm{bolo},i})}
{P_{\mathrm{bolo},i}\ln 10} \right)^2 + \left(\sigma_{sys}^{(\alpha)}\right)^2, \nonumber \\
& & \mathrm{for}\,\, \alpha=1,2,3,5 \\
\sigma_{i,4}^2 &=& \sigma_{a_4}^2+ \left( \sigma_{b_4} 
x_i^{(4)}\right)^2 + \left( \frac{b_4 \sigma(x_{0,i}^{(4)})}
{x_{0,i}^{(4)}\ln 10} \right)^2  \nonumber\\
& & + \left( \frac{\sigma(S_{\mathrm{bolo},i})}{S_{\mathrm{bolo},i}\ln 10} \right)^2
+\left( \frac{\sigma(F_{\mathrm{beam},i})}{F_{\mathrm{beam},i}\ln 10} \right)^2
\nonumber\\
& & + \left(\sigma_{sys}^{(4)}\right)^2
\ea

For each cosmological model given by $\overline{d_L}^2$, we can calibrate
the GRBs as described in Sec.\ref{sec:cali}, and then derive the distance
estimate $\left(\log \overline{d_L}^2\right)_i$ from each GRB.
The $\chi^2$ of a model is given by
\be
\chi^2_{GRB}= \sum_{i=1}^{N_{GRB}} \frac{ \left[ \left(\log \overline{d_L}^2\right)_i^{data}-
\log \overline{d_L}^2(z_i) \right]^2}
{ \left[\sigma\left(\log \overline{d_L}^2\right)_i^{data}\right]^2}.
\label{eq:chi2GRB}
\ee
The treatment of the asymmetric errors in $E_{\mathrm{peak}}$ is given
by Eq.(\ref{eq:dx asy}).

We summarize the cosmological constraints from GRB data, 
$\overline{d_L}^2(z_{GRB})$, in terms of a set of model-independent
distance measurements $\{\overline{r_p}(z_i)\}$:
\be
\label{eq:rp}
\overline{r_p}(z_i)\equiv \frac{r_p(z)}{r_p(0.17)}, \hskip 1cm
r_p(z) \equiv \frac{(1+z)^{1/2}}{z}\, \frac{H_0}{ch}\, r(z),
\ee
where $r(z)=d_L(z)/(1+z)$ is the comoving distance at $z$.
For the 69 GRBs from \cite{Schaefer07}, the lowest redshift GRB has
$z=0.17$, while the highest redshift GRB has $z=6.6$.
There are only four GRBs at $4.5 \leq z \leq 6.6$.
We find that the optimal binning is to divide the redshift range
between 0.17 and 4.5 into 5 bins, and choosing the last bin to
span from 4.5 and 6.6 (see Fig.2).

Note that the ratio $r_p(z)/r_p(0.17)$ is the most convenient 
distance parameter choice for the currently available GRB data, 
since $z=0.17$ is the lowest redshift
GRB in the data set, and the absolute calibration of GRBs is unknown.
Using the distance ratio $r_p(z)/r_p(0.17)$ removes the dependence 
on Hubble constant (which is
unknown due to the unknown absolute calibration of GRBs).

\begin{figure} 
\psfig{file=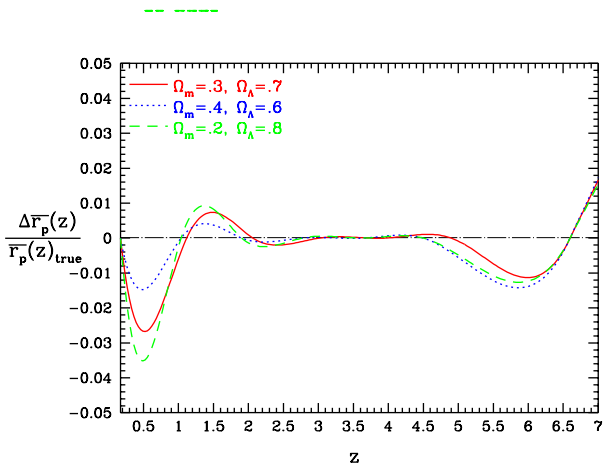,width=4in}\\
\caption{\label{fig:rpa}\footnotesize%
The accuracy with which the scaled distance $\overline{r_p}(z)=r_p(z)/r_p(0.17)$
can be reconstructed using cubic spline interpolation (C.S.)
from $\{\overline{r_p}(z_i)\}$ for $N_{bin}=6$ with our choice of binning.
Note that $\Delta \overline{r_p}(z)=\overline{r_p}(z)_{C.S.}-\overline{r_p}(z)_{true}$.}
\end{figure}

Because $\{\overline{r_p}(z_i)\}$ varies very slowly for all
cosmological models allowed by current data, the scaled
distance $\overline{r_p}(z)$ at an arbitary
redshift $z$ can be found using cubic spline interpolation
from $\{\overline{r_p}(z_i)\}$ to $\sim$ 1-3\% percent accuracy
for $N_{bin}=6$ with our choice of binning, see Fig.\ref{fig:rpa}.

Note that for a given set of possible values of $\{\overline{r_p}(z_i)\}$ 
($i=1,2,...,6$), the luminosity distance at an arbitrary redshift, $d_L(z)$, 
is given by the accurate interpolation described above.
Thus {\it no} assumptions about cosmological parameters are made.
We calibrate the GRBs for each set of possible values of $\{\overline{r_p}(z_i)\}$ 
($i=1,2,...,6$), and compute the likelihood of this set of
$\{\overline{r_p}(z_i)\}$ in a Markov Chain Monte Carlo analysis.
Hence the distances $\{\overline{r_p}(z_i)\}$ are {\it independent}
of assumptions about cosmological parameters.

\subsection{Other cosmological data}
\label{sec:other data}

GRB data alone do not constrain dark energy parameters.
In order to investigate how well our model-independent distance
measurements from GRB represent GRB data, we study them
in combination with 307 SNe Ia \cite{Kowa08}, 
Cosmic Microwave Background anisotropy (CMB) data 
from WMAP five year observations \cite{Komatsu08},
and baryon acoustic oscillation (BAO) scale measurement
from the SDSS data \cite{Eisen05}.

SN Ia data give the luminosity distance as a function of redshift,
$d_L(z)=(1+z)\, r(z)$.
We use 307 SNe Ia from the ``Union'' compilation by
Ref.\cite{Kowa08}, which includes data from Ref.\cite{SNdata}.
Appendix A of Ref.\cite{WangPia07} describes in detail how we use 
SN Ia data (flux-averaged to reduce lensing-like systematic 
effects \cite{flux-avg} and marginalized over $H_0$) in 
this paper.\footnote{A public code for flux-averaging
SN Ia data is available at http://www.nhn.ou.edu/$\sim$wang/SNcode/.}
We applied flux-averaging to the ``without systematics'' data from 
the ``Union'' compilation. The ``with systematics'' data
differ only in having small offsets added to
different data sets, which leads to correlations in the data.
We expect the increase of uncertainties resulting from flux-averaging
to be larger than the effect of the small offsets between
different data sets.

We include the CMB data using the method proposed
by Ref.\cite{WangPia07}, which showed that
the CMB shift parameters
\be
R \equiv \sqrt{\Omega_m H_0^2} \,r(z_*), \hskip 0.1in
l_a \equiv \pi r(z_*)/r_s(z_*),
\ee
together with $\Omega_b h^2$, provide an efficient summary
of CMB data as far as dark energy constraints go.
We use the covariance matrix of [$R(z_*), l_a(z_*), \Omega_b h^2]$ from
the five year WMAP data (Table 11 of \cite{Komatsu08}), with $z_*$
given by fitting formulae from Hu \& Sugiyama (1996) \cite{Hu96}.
CMB data are included in our analysis by adding
the following term to the $\chi^2$ of a given model
with $p_1=R(z_*)$, $p_2=l_a(z_*)$, and $p_3=\Omega_b h^2$:
\be
\label{eq:chi2CMB}
\chi^2_{CMB}=\Delta p_i \left[ \mathrm{Cov}_{CMB}^{-1}(p_i,p_j)\right]
\Delta p_j,
\hskip .5cm
\Delta p_i= p_i - p_i^{data},
\ee
where $p_i^{data}$ are the maximum likelyhood values given in 
Table 10 of \cite{Komatsu08}.


We also use the SDSS baryon acoustic oscillation (BAO)
scale measurement by adding the following term to the
$\chi^2$ of a model:
\be
\chi^2_{BAO}=\left(\frac{A-A_{BAO}}{\sigma_A}\right)^2,
\label{eq:chi2bao}
\ee
where $A$ is defined as
\be
\label{eq:A}
A = \left[ r^2(z_{BAO})\, \frac{cz_{BAO}}{H(z_{BAO})} \right]^{1/3} \, 
\frac{\left(\Omega_m H_0^2\right)^{1/2}} {cz_{BAO} },
\ee
and $A_{BAO}=0.469\,(n_S/0.98)^{-0.35}$,
$\sigma_A= 0.017$, and $z_{BAO}=0.35$
(independent of a dark energy model) \cite{Eisen05}. 
We take the scalar spectral index $n_S=0.96$ as measured by WMAP
five year observations \cite{Komatsu08}.

Finally, in combination with CMB data, we include the Hubble Space Telescope
(HST) prior on the Hubble constant of $h=0.72\pm 0.08$ \cite{HST_H0}.

\section{Results}

For Gaussian distributed measurements, the likelihood function
$L\propto e^{-\chi^2/2}$, with 
\be
\chi^2=\chi^2_{GRB}+\chi^2_{SNe}+\chi^2_{CMB}+\chi^2_{BAO},
\label{eq:chi2}
\ee
where $\chi^2_{GRB}$ is given in Eqs.({\ref{eq:rGRB1}})-({\ref{eq:rGRB3}}),
$\chi^2_{SNe}$ is given in Appendix A of Ref.\cite{WangPia07},
$\chi^2_{CMB}$ is given in Eq.({\ref{eq:chi2CMB}}),
and $\chi^2_{BAO}$ is given in Eq.({\ref{eq:chi2bao}}).

We run a Monte Carlo Markov Chain (MCMC) based on the MCMC engine 
of \cite{Lewis02} to obtain ${\cal O}$($10^6$) samples for each set of 
results presented in this paper. 
For the model-independent GRB distance measurements, the parameter
set is $\{\overline{r_p}(z_i)\}$ ($i=1,2,...,6$); {\it no} assumptions
are made about cosmological parameters.
For the combined analysis of GRBs (using either the model-independent
distance measurements $\{\overline{r_p}(z_i)\}$ ($i=1,2,...,6$),
or the 69 GRBs directly) with other cosmological 
data, the cosmological parameter sets used are:
$\Omega_m$ for a flat Universe with a cosmological constant;
($\Omega_m$, $\Omega_\Lambda$) for a cosmological
constant; ($\Omega_m$, $w_0$) for a flat Universe with a constant dark 
eneagy equation of state; and ($\Omega_m$, $h$, $\Omega_b h^2$, $p_{\mathrm{DE}}$)
for a flat Universe with a dark eneagy equation of state linear in
the cosmic scale factor, with $p_{\mathrm{DE}}=(w_0$, $w_{0.5}$) or ($w_0$, $w_a$).
We assum flat priors for all the parameters, and allow ranges 
of the parameters wide enough such that further increasing the allowed 
ranges has no impact on the results.
The chains typically have worst e-values (the
variance(mean)/mean(variance) of 1/2 chains)
much smaller than 0.005, indicating convergence.
The chains are subsequently 
appropriately thinned to ensure independent samples.

\begin{figure} 
\psfig{file=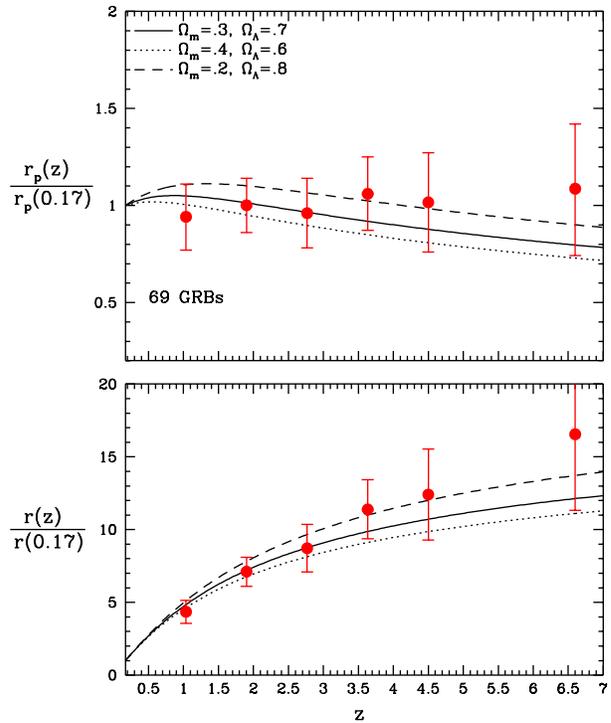,width=4in}\\
\caption[2]{\label{fig:rp}\footnotesize%
The distances measured from 69 GRBs using the five calibration relations in Eqs.(\ref{eq:C1})-(\ref{eq:C5}).
The error bars indicate the 68\% C.L. uncertainties.
The top panel shows the scaled distances $r_p(z_i)$ (see Eq.(\ref{eq:rp}));
the bottom panel shows the corresponding comoving distances $r(z_i)$.}
\end{figure}

Fig.\ref{fig:rp} shows the distances $\{\overline{r_p}(z_i)\}$
measured from 69 GRBs using the five calibration relations 
in Eqs.(\ref{eq:C1})-(\ref{eq:C5}).
Table 2 gives the mean and 68\% confidence level errors of
$\{\overline{r_p}(z_i)\}$. The normalized covariance matrix of 
$\{\overline{r_p}(z_i)\}$ is given in Table 3.
To use our GRB distance measurements to constrain cosmological models, use
\ba
\label{eq:rGRB1}
\chi^2_{GRB}&=& \left[\Delta \overline{r_p}(z_i)\right]  \cdot
\left(\mathrm{Cov}^{-1}_{GRB}\right)_{ij}\cdot
\left[\Delta \overline{r_p}(z_j)\right]
\nonumber\\
\Delta \overline{r_p}(z_i)&=& \overline{r_p}^{\mathrm{data}}(z_i)-\overline{r_p}(z_i),
\ea
where $\overline{r_p}(z)$ is defined by Eq.(\ref{eq:rp}).
The covariance matrix is given by
\be
\left(\mathrm{Cov}_{GRB}\right)_{ij}=
\sigma(\overline{r_p}(z_i)) \sigma(\overline{r_p}(z_j)) 
\left(\overline{\mathrm{Cov}}_{GRB}\right)_{ij},
\ee
where $\overline{\mathrm{Cov}}_{GRB}$ is the normalized covariance matrix
from Table 3, and
\ba
\label{eq:rGRB3}
\sigma(\overline{r_p}(z_i)) &=&\sigma\left(\overline{r_p}(z_i)\right)^+, \hskip 0.5cm \mathrm{if}\,\, 
\overline{r_p}(z) \ge \overline{r_p}(z)^{\mathrm{data}}; \nonumber\\
\sigma(\overline{r_p}(z_i)) &=&\sigma\left(\overline{r_p}(z_i)\right)^-, \hskip 0.5cm \mathrm{if}\,\, 
\overline{r_p}(z) < \overline{r_p}(z)^{\mathrm{data}},
\ea
where $\sigma\left(\overline{r_p}(z_i)\right)^+$ and 
$\sigma\left(\overline{r_p}(z_i)\right)^-$ are the 68\% C.L. errors
given in Table 2.

\begin{table*}[htb]
\caption{Distances measured from 69 GRBs with 68\% C.L. upper and lower uncertainties.}
\begin{center}
\begin{tabular}{|l|llll|}
\hline
 & $z$ & $\overline{r_p}^{data}(z)$ & $\sigma\left(\overline{r_p}(z)\right)^+$ & 
 $\sigma\left(\overline{r_p}(z)\right)^-$\\
 \hline
    0 &     0.17    &    1.0000     &     --     &   -- \\
    1 &     1.036  &    0.9416   &   0.1688   &   0.1710\\
    2 &     1.902  &    1.0011   &   0.1395   &   0.1409\\
    3 &     2.768  &    0.9604   &   0.1801   &   0.1785\\
    4 &     3.634  &    1.0598   &   0.1907   &   0.1882\\
    5 &     4.500  &    1.0163   &   0.2555    &  0.2559\\
    6 &     6.600  &    1.0862   &   0.3339    &  0.3434 \\
\hline
\end{tabular}
\end{center}
\end{table*}

\begin{table*}[htb]
\caption{Normalized covariance matrix of distances measured from 69 GRBs}
\begin{center}
\begin{tabular}{|l|llllll|}
\hline
      & 1 & 2 & 3 &4 &5 &6\\
      \hline
1 &  1.0000 & 0.7056  & 0.7965  & 0.6928 &  0.5941  & 0.5169 \\
2 &  0.7056  & 1.0000 & 0.5653  & 0.6449  & 0.4601  & 0.4376 \\
3 & 0.7965  & 0.5653  & 1.0000 & 0.5521  & 0.5526  & 0.4153 \\
4 & 0.6928  & 0.6449  & 0.5521  & 1.0000 & 0.4271  & 0.4242 \\
5 & 0.5941  & 0.4601  & 0.5526  & 0.4271  & 1.0000 & 0.2999 \\
6 & 0.5169  & 0.4376  & 0.4153  & 0.4242  & 0.2999  & 1.0000\\
\hline
\end{tabular}
\end{center}
\end{table*}


Using the distance measurements from GRBs (see Tables 2 and 3),
we find $\Omega_m=$0.247 (0.122, 0.372) (mean and 68\% C.L. range).
Using the 69 GRBs directly, we find
$\Omega_m=$0.251 (0.135, 0.365). 
This demonstrates that our
model-independent distance measurements from GRBs can be used
as a useful summary of the current GRB data.

\begin{figure} 
\psfig{file=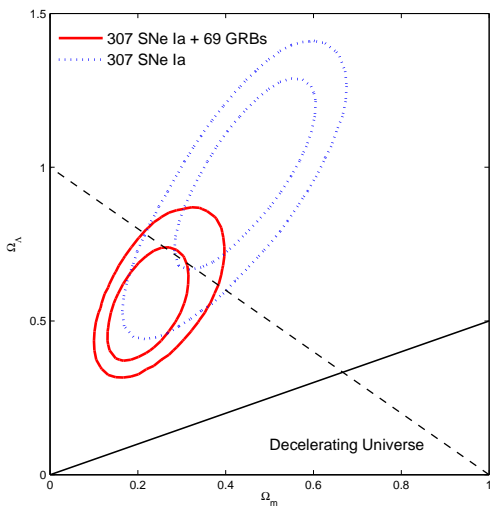,width=2.6in}\\
\caption[2]{\label{fig:OmOX}\footnotesize%
The joint confidence contours for ($\Omega_m$, $\Omega_\Lambda$), 
from an analysis of 307 SNe Ia with and without 69 GRBs.
A cosmological constant is assumed.}
\end{figure}

\begin{figure} 
\psfig{file=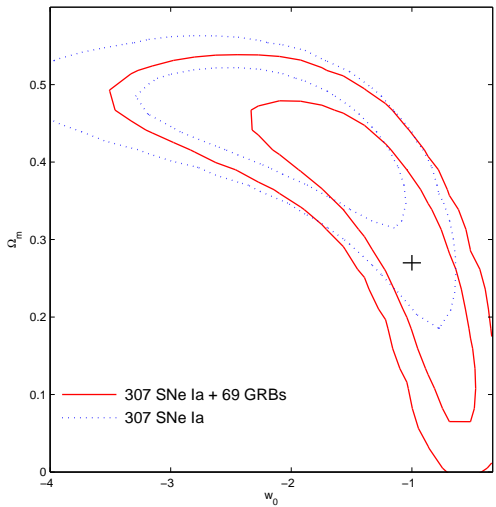,width=2.6in}\\
\caption[2]{\label{fig:wconst}\footnotesize%
The joint confidence contours for ($w_0$, $\Omega_m$), 
from an analysis of 307 SNe Ia with and without 69 GRBs.
A flat Universe and dark energy with constant equation of
state are assumed.}
\end{figure}

Fig.\ref{fig:OmOX} shows the joint confidence contours for 
($\Omega_m$, $\Omega_\Lambda$), from an analysis of 307 SNe Ia with and without 69 GRBs,
assuming a cosmological constant. This shows that the addition of GRB
data significantly reduces the uncertainties in ($\Omega_m$, $\Omega_\Lambda$),
and shifts the bestfit parameter values towards a lower matter density Universe.
Fig.\ref{fig:wconst} shows the joint confidence contours for ($w_0$, $\Omega_m$), 
from an analysis of 307 SNe Ia with and without 69 GRBs,
assuming a flat Universe. This shows that SNe Ia alone rules out
a cosmological constant at greater than 68\% C.L., but the addition
of GRB data significantly shifts the bestfit parameter values, and
a cosmological constant is consistent with combined SN Ia and GRB data
at 68\% C.L.

\begin{figure} 
\psfig{file=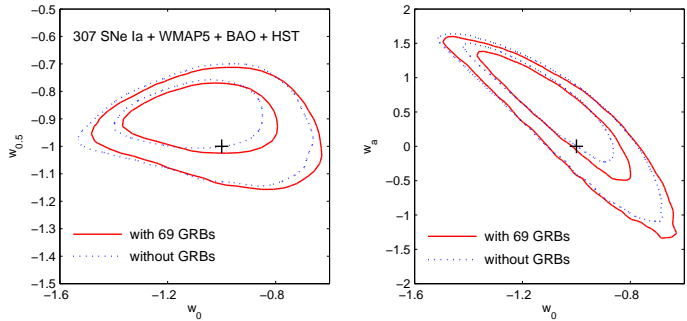,width=3.6in}\\
\caption[2]{\label{fig:w0wa}\footnotesize%
The joint confidence contours for ($w_0$, $w_{0.5}$) and ($w_0$, $w_a$), 
from a joint analysis of 307 SNe Ia with CMB data from WMAP5, 
and SDSS BAO scale measurement, with and without 69 GRBs.
HST prior on $H_0$ has been imposed and 
a flat Universe is assumed.}
\end{figure}

We now consider a dark energy equation of state linear in the cosmic scale 
factor $a$. Fig.\ref{fig:w0wa} shows the joint confidence contours for 
($w_0$, $w_{0.5}$) and ($w_0$, $w_a$), 
from a joint analysis of 307 SNe Ia with CMB data from WMAP5, 
and SDSS BAO scale measurement, with and without 69 GRBs.
HST prior on $H_0$ has been imposed and 
a flat Universe is assumed.
Note that $w_{0.5}=w_X(z=0.5)$ in the linear parametrization \cite{Wang08}
\ba
w_X(a)&=&\left(\frac{a_c-a}{a_c-1}\right) w_0
+\left(\frac{a-1}{a_c-1}\right) w_{0.5} \nonumber\\
&=&\frac{a_cw_0-w_{0.5} + a(w_{0.5}-w_0)}{a_c-1}
\label{eq:wc}
\ea
with $a_c=2/3$ (i.e., $z_c=0.5$). Eq.(\ref{eq:wc}) corresponds to a 
dark energy density function
\ba
X(z) =\frac{\rho_X(z)}{\rho_X(0)}&=&\exp 
\left\{ 3\left[ 1+\left(\frac{a_cw_0-w_{0.5}}{a_c-1}\right)\right]\,\ln(1+z)
\right.\nonumber\\
& & + \left.3 \left(\frac{w_{0.5}-w_0}{a_c-1}\right) \frac{z}{1+z}\right\}
\ea
Eq.(\ref{eq:wc}) is related to $w_X(z)=w_0+(1-a)w_a$ \cite{Chev01} 
by setting \cite{Wang08}
\be
w_a= \frac{w_{0.5}-w_0}{1-a_c}, \hskip 1cm
{\rm or} \hskip 1cm
w_{0.5}=w_0+(1-a_c) w_a.
\label{eq:wa,wc}
\ee
Ref.\cite{Wang08} showed that ($w_0$, $w_{0.5}$) are much less
correlated than ($w0$, $w_a$), thus are a better set of parameters
to use. Fig.{\ref{fig:w0wa}} shows that the addition of
GRB data notably shifts the 68\% C.L. contours of ($w_0$,$w_{0.5}$) 
and ($w_0$, $w_a$) to enclose the cosmological constant model
($w_X(a)=-1$).

\section{Summary and Discussion}

We have shown that the current GRB data, consisting of 69 GRBs
spanning the redshift range from 0.17 to 6.6, can be
summarized by a set of distance measurements (see Fig.\ref{fig:rp}).
For each set of possible distance values, the GRBs are calibrated.
The resultant distance measurements (given in Tables 2 and 3) 
are independent of cosmology,
and can be easily used to combine with other cosmological data sets
to constrain dark energy (see Eqs.[\ref{eq:rGRB1}]-[\ref{eq:rGRB3}] 
and [\ref{eq:rp}]).

The number of bins used in our distance measurement, $n_{bin}=6$, 
is determined by the current sample of GRBs. Increasing the number 
of bins by one leads to oscillations in the measured distance values.
As more GRB data become available, we can expect to be able to
increase the number of bins used to represent GRB data.

We find that GRB data alone give $\Omega_m=0.25^{+0.12}_{-0.11}$
for a flat Universe with a cosmological constant.
Fig.{\ref{fig:OmOX}} and Fig.{\ref{fig:wconst}} show that
combining GRB data with SN Ia data significantly shifts the
bestfit model towards a lower matter density Universe, in agreement
with galaxy redshift survey \cite{Tegmark06} and CMB data \cite{Dunkley08}.
Fig.\ref{fig:w0wa} shows that assuming a dark energy equation of state 
linear in cosmic scale factor $a$, including GRB data together
with SN Ia, CMB, and BAO data shifts the 68\% C.L. contours
of the two dark energy parameters to enclose the cosmological
constant model.

For a flat Universe with a cosmological constant,
our results for GRBs alone differ from that of Ref.\cite{Schaefer07}.
We find $\Omega_m=0.25^{+0.12}_{-0.11}$, while
Ref.\cite{Schaefer07} found $\Omega_m=0.39^{+0.12}_{-0.08}$.
The difference likely resulted from the different numerical methods
used to calibrate the GRBs. We have fitted straight lines
to the calibration relations using errors in both $x$ and $y$
coordinates (see Eq.[\ref{eq:chi2fit}]), while Ref.\cite{Schaefer07} 
used ``ordinary least squares without any weighting''.
Our results are consistent with the results of
Ref.\cite{Liang08}, who used SNe Ia to help calibrate the GRBs
and found $\Omega_m=0.25_{-0.05}^{+0.04}$.

In our quest to solve the mystery of dark energy, GRBs will
provide a unique and complementary probe. 
Our results will make it very convenient to incorporate GRB data 
in any joint cosmological data analysis in a simple and robust manner.

\bigskip

{\bf Acknowledgements}
I acknowledge the use of getdist from cosmomc
in processing the MCMC chains.

\end{document}